# We are IntechOpen,
# the world's leading publisher of Open Access books
# Built by scientists, for scientists

## 4,400
Open access books available

## 117,000
International authors and editors

## 130M
Downloads

## 154
Countries delivered to

Our authors are among the

## TOP 1%
most cited scientists

## 12.2%
Contributors from top 500 universities

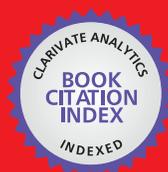

**WEB OF SCIENCE**™

Selection of our books indexed in the Book Citation Index
in Web of Science™ Core Collection (BKCI)

# Interested in publishing with us?
# Contact book.department@intechopen.com

Numbers displayed above are based on latest data collected.
For more information visit www.intechopen.com

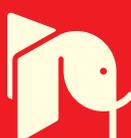

Chapter

# Secure State Estimation and Attack Reconstruction in Cyber-Physical Systems: Sliding Mode Observer Approach

*Shamila Nateghi, Yuri Shtessel, Christopher Edwards and Jean-Pierre Barbot*

**Abstract**

A cyber-physical system (CPS) is a tight coupling of computational resources, network communication, and physical processes. They are composed of a set of networked components, including sensors, actuators, control processing units, and communication agents that instrument the physical world to make "smarter." However, cyber components are also the source of new, unprecedented vulnerabilities to malicious attacks. In order to protect a CPS from attacks, three security levels of protection, detection, and identification are considered. In this chapter, we will discuss the identification level, i.e., secure state estimation and attack reconstruction of CPS with corrupted states and measurements. Considering different attack plans that may assault the states, sensors, or both of them, different online attack reconstruction approaches are discussed. Fixed-gain and adaptive-gain finite-time convergent observation algorithms, specifically sliding mode observers, are applied to online reconstruction of sensor and state attacks. Next, the corrupted measurements and states are to be cleaned up online in order to stop the attack propagation to the CPS via the control signal. The proposed methodologies are applied to an electric power network, whose states and sensors are under attack. Simulation results illustrate the efficacy of the proposed observers.

## 1. Introduction

Cyber-physical systems (CPS) are the integration of the cyber-world of computing and communications with the physical world. In many systems, control of a physical plant is integrated with a wireless communication network, for example, transportation networks, electric power networks, integrated biological systems, industrial automation systems, and economic systems [1, 2]. Since CPSs use open computation and communication platform architectures, they are vulnerable to suffering adversarial physical faults or cyber-attacks. Faults and cyber-attacks are referred to as *attacks* throughout this chapter.

Recent real-world cyber-attacks, including multiple power blackouts in Brazil [3], and the Stuxnet attack [4] in 2010, showed the importance of providing security to CPSs. Identification and modeling process as [5, 6] which are based on data can be seriously affected by corrupted data. As a result, information security techniques [7] may be not sufficient for protecting systems from sophisticated cyber-attacks. It is suggested in [8] that information security mechanisms have to be complemented by specially designed resilient control systems. Controlling CPS with sensors and actuators, who are hijacked/corrupted remotely or physically by the attackers, is a challenge. The use of novel control/observation algorithms is proposed in this chapter for recovering CPS performance online if an attacker penetrates the information security mechanisms.

Cyber security of CPS must provide three main security goals: *availability*, *confidentiality*, and *integrity* [7]. This means that the CPS is to be accessible and usable upon demand, the information has to be kept secret from unauthorized users, and the trustworthiness of data has to be guaranteed. Lack of availability, confidentiality, and integrity yields denial of service, disclosure, and deception, respectively. A specific kind of deception attack called a *replay attack* has been investigated when the system model is unknown to the attackers but they have access to the all sensors [9, 10]. *Replay attacks* are carried out by "hijacking" the sensors, recording the readings for a certain time, and repeating such readings while injecting them together with an exogenous signal into the system's sensors. It is shown that these attacks can be detected by injecting a random signal, unknown to the attacker, into the system. In the case when the system's dynamic model is known to the attacker, another kind of deception attack, called a *cover attack*, has been studied in [11], and the proposed algorithm allows cancelling out the effect of this attack on the system dynamics. In systems with unstable modes, false data injection attacks are applied to make some unstable modes unobservable [12]. Denial of service attacks assaults data availability through blocking information flows between different components of the CPS. The attacker can jam the communication channels, modify devices, and prevent them from sending data, violate the routing protocols, etc. [13]. In a stealth attack, the attacker modifies some sensor readings by physically tampering with the individual meters or by getting access to some communication channels [14, 15]. As a result, detecting and isolating of cyber-attacks in CPSs has received immense attention [16]. However, how to ensure the CPS can continue functioning properly if a cyber-attack has happened is another serious problem that should be investigated; therefore, the focus of this chapter is on resilient control of CPS.

In [17], new adaptive control architectures that can foil malicious sensor and actuator attacks are developed without reconstructing the attacks, by means of feedback control only. A sparse recovery algorithm is applied to reconstruct online the cyber-attacks in [18]. Sliding mode control with advantages of quick response and strong robustness is one of the best approaches to control CPS [19–22]. In [23], a finite-time convergent higher-order sliding mode (HOSM) observer, based on a HOSM differentiator and a sparse recovery algorithm, are used to reconstruct online the cyber-attack in a nonlinear system. Detection and observation of a scalar attack by a sliding mode observer (SMO) has been accomplished for a linearized differential-algebraic model of an electric power network when plant and sensor attacks do not occur simultaneously [24]. Cyber-attacks against phasor measurement unit (PMU) networks are considered in [25], where a risk mitigation technique determines whether a certain PMU should be kept connected to network or removed. In [26] a sliding mode-based observation algorithm is used to reconstruct the attacks asymptotically. This reconstruction is approximate only, since pseudo-inverse techniques are used.

In this chapter, CPSs controlled by a control input subject to sensor attacks and state/plant attacks are considered. The corrupted measurements propagate the attack signals to the CPS through the control signals causing CPS performance degradation. The main challenge that is addressed in the chapter is online exact reconstruction of the sensor and state attacks with an application to an electric power network. The contribution of this chapter is:

- Novel fixed and adaptive-gain SMO for the linearized/linear CPS under attack are proposed for the online reconstruction of sensor attacks. The *time-varying* attacks are reconstructed via the proposed SMO that includes a newly designed dynamic filter. Note that the well-known SMO proposed in [27] reconstructs the slow-varying perturbations only.

- A super twisting SMO is applied to reconstruct the state/plant time-varying attacks of the linearized/linear CPS under attack.

- For online state/plant attack reconstruction in *nonlinear* CPS under attack, a higher-order sliding mode disturbance observer [28] is used.

- An algorithm that use sliding mode differentiation techniques [29] in concert with the finite-time convergent observer for the sparse signal recovery is applied to online reconstruction of time-varying attack in nonlinear CPS under attack when we have limited measurements and more possible sources of attack [30].

## 2. Motivation example: electric power network under attack

In a real-world power network, only a small group of generator rotor angles and rates is directly measured, and typical attacks aim at injecting disturbance signals that mainly affect the sensorless generators [24].

The small-signal version of the classic structure-preserving power network model is adopted to describe the dynamics of a power network. Consider a connected power network consisting of $n_1$ generators $\{g_1, ..., g_{n_1}\}$ and $n_2$ load buses $\{b_{n_1+1}, ..., b_{n_1+n_2}\}$. The interconnection structure of the power network is encoded by a connected susceptance-weighted graph $G$. The vertices of $G$ are the generators $g_i$ and the buses $b_i$. The edges of $G$ are the transmission lines $\{b_i, b_j\}$ and the connections $\{g_i, b_i\}$ weighted by their susceptance values. The Laplacian associated with the susceptance-weighted graph is the symmetric susceptance matrix $L \in \mathbb{R}^{(n_1+n_2) \times (n_1+n_2)}$ defined by $L^\theta = \begin{bmatrix} L^\theta_{g,g} & L^\theta_{g,l} \\ L^\theta_{l,g} & L^\theta_{l,l} \end{bmatrix}$ [8].

The CPS that motivates the results presented in this work is the US Western Electricity Coordinating Council (WECC) power system [8] under attack with three generators and six buses, whose electrical schematic is presented in **Figure 1**. The mathematical model of the power network in **Figure 1** under sensor stealth attack and deception attack can be represented as the following descriptor equations that consist of differential and algebraic equations [8]:

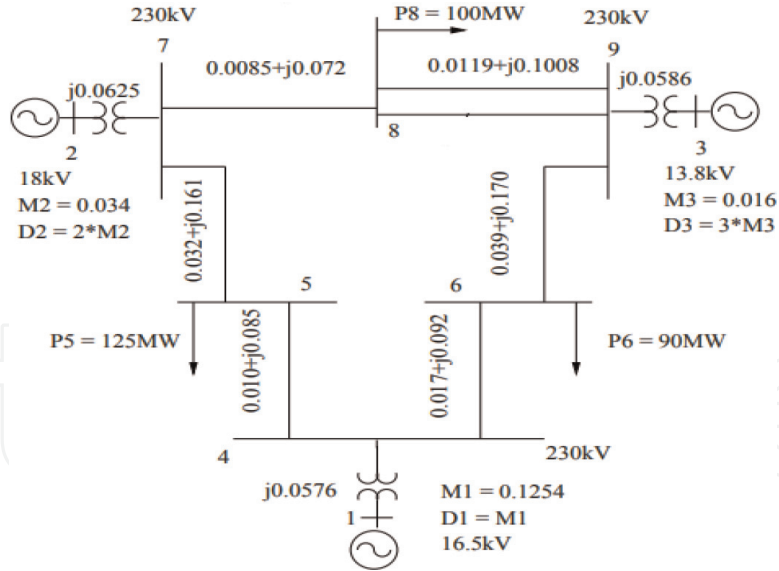

**Figure 1.**
*The WECC power system [8].*

$$\begin{bmatrix} I & 0 & 0 \\ 0 & M_g & 0 \\ 0 & 0 & 0 \end{bmatrix} \begin{bmatrix} \dot{\delta} \\ \dot{\omega} \\ \dot{\theta} \end{bmatrix} = -\begin{bmatrix} 0 & -I & 0 \\ L_{g,g}^\theta & E_g & L_{g,l}^\theta \\ L_{l,g}^\theta & 0 & L_{l,l}^\theta \end{bmatrix} \underbrace{\begin{bmatrix} \delta \\ \omega \\ \theta \end{bmatrix}}_{x} + \underbrace{\begin{bmatrix} 0 \\ B_\omega \\ B_\theta \end{bmatrix}}_{B} d_x + \begin{bmatrix} 0 \\ P_\omega \\ P_\theta \end{bmatrix}, \quad y = Cx + Dd_y \quad (1)$$

where the state vector $x = \begin{bmatrix} \delta^T & \omega^T & \theta^T \end{bmatrix}^T$ includes the vector of rotor angles $\delta \in \mathbb{R}^3$, the vector of generator speed deviations from synchronicity $\omega \in \mathbb{R}^3$, as well as the vector of voltage angles at the buses $\theta \in \mathbb{R}^6$. The $y \in \mathbb{R}^p$ is the measurement vector, $d_x \in \mathbb{R}^{m_1}$ is the *Deception* attack corrupting the states, and $d_y \in \mathbb{R}^{m-m_1}$ is the *stealth* attack vector spoofing the measurements. Note that the states of the plant are under attack even if they are not attacked directly but via propagation.

The measurement corruption attacks through an output control feedback. The matrices $E_g, M_g \in \mathbb{R}^{3\times 3}$ are diagonal whose nonzero entries consist of the damping coefficients and the normalized inertias of the generators, respectively:

$$M_g = \begin{bmatrix} 0.125 & 0 & 0 \\ 0 & 0.034 & 0 \\ 0 & 0 & 0.016 \end{bmatrix}, \quad E_g = \begin{bmatrix} 0.125 & 0 & 0 \\ 0 & 0.068 & 0 \\ 0 & 0 & 0.048 \end{bmatrix} \quad (2)$$

The inputs $P_\omega$ and $P_\theta$ are due to *known* changes in the mechanical input power to the generators and real power demands at the loads. The matrices $B \in \mathbb{R}^{12\times m_1}$ and $D \in \mathbb{R}^{p\times (m-m_1)}$ are the attack distribution matrices, and $C \in \mathbb{R}^{p\times 12}$ is the output gain matrix. The $L^\theta \in \mathbb{R}^{9\times 9}$ with $L_{g,g}^\theta \in \mathbb{R}^{3\times 3}$, $L_{g,l}^\theta \in \mathbb{R}^{3\times 6}$, $L_{l,g}^\theta \in \mathbb{R}^{6\times 3}$, $L_{l,l}^\theta \in \mathbb{R}^{6\times 6}$ is giving by

$$L^\theta = \begin{bmatrix}
0.058 & 0 & 0 & -0.058 & 0 & 0 & 0 & 0 & 0 \\
0 & 0.063 & 0 & 0 & -0.063 & 0 & 0 & 0 & 0 \\
0 & 0 & 0.059 & 0 & 0 & -0.059 & 0 & 0 & 0 \\
-0.058 & 0 & 0 & 0.265 & 0 & 0 & -0.085 & -0.092 & 0 \\
0 & -0.063 & 0 & 0 & 0.296 & 0 & -0.161 & 0 & -0.072 \\
0 & 0 & -0.059 & 0 & 0 & 0.330 & 0 & -0.170 & -0.101 \\
0 & 0 & 0 & -0.085 & -0.161 & 0 & 0.246 & 0 & 0 \\
0 & 0 & 0 & -0.092 & 0 & -0.170 & 0 & 0.262 & 0 \\
0 & 0 & 0 & 0 & -0.072 & -0.101 & 0 & 0 & 0.173
\end{bmatrix} \quad (3)$$

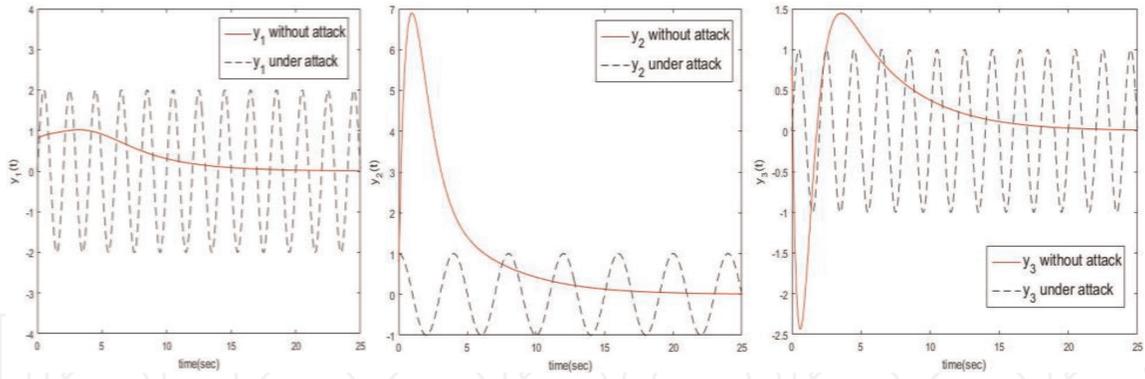

**Figure 2.**
*Comparing corrupted sensor measurements ($\omega_1, \omega_2, \omega_3$ under attack) and sensor measurements when there is no attack.*

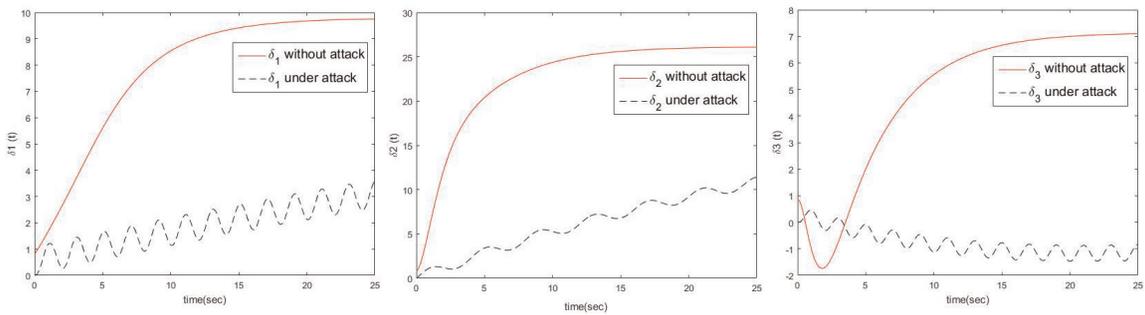

**Figure 3.**
*Comparing corrupted states ($\delta_1, \delta_2, \delta_3$ under attack) and stats when there is no attack.*

Note that $\omega_i \to 0 \ \forall i = 1, 2, 3$ in a case of the nominal performance of the studied network. Consider the case when the outputs of system, which are the measurement sensors $\omega_1, \omega_2, \omega_3$, are corrupted by the following *stealth* attacks.

$$d_1 = -\omega_1 + 2\sin(\pi t), \quad d_2 = -\omega_2 + \cos(0.5\pi t), \quad d_3 = -\omega_3 + \sin(\pi t) \quad (4)$$

The system (1) was simulated with and without above attacks. Based on the simulation results shown in **Figures 2** and **3**, the stealth attack in (4) yields inappropriate degradation of the power network performance.

This motivates why online reconstruction of the attacks followed by cleanup of the measurements prior to using them in control signal is of prime importance for retaining the performance of the power network (as it will be shown in Section VI where the proposed SMO is applied to achieve this goal). The case study of the power network (1) will be further discussed in details in Section 6.

## 3. Cyber-physical system dynamics

Consider the following completely observable and asymptotically stable system

$$\begin{aligned} \dot{x} &= f(x) + B(x)d(t) \\ y &= C(x) + Dd(t) \end{aligned} \quad (5)$$

where $x \in \mathbb{R}^n$ is the state vector, $f(x) \in \mathbb{R}^n$ is a smooth vector field, $d(t) \in \mathbb{R}^m$ denotes the attack/fault vector which is additive and matched to the control signal, $y \in \mathbb{R}^p$ is the measurement vector, $p \geq m$, $C(x) \in \mathbb{R}^p$ is the output smooth vector field, $B(x) \in \mathbb{R}^{n \times m}$ and $D \in \mathbb{R}^{p \times m}$ denote the attack/fault distribution matrices. For notational convenience, and without affecting generality, the input distribution matrices can be partitioned as

$$B(x) = [B_1(x) \quad \mathbf{0}_1], D = [\mathbf{0}_2 \quad D_1] \tag{6}$$

where $B_1(x) \in \mathbb{R}^{n \times m_1}$, $D_1 \in \mathbb{R}^{p \times (m-m_1)}$, $\mathbf{0}_1 \in \mathbb{R}^{n \times (m-m_1)}$, $\mathbf{0}_2 \in \mathbb{R}^{p \times m_1}$ where $m_1 \leq m$.

**Assumption (A1):** $B_1(x), D_1$ are of full rank.

The attack/fault vector is partitioned accordingly as

$$d = \begin{bmatrix} d_x \\ d_y \end{bmatrix} \quad \text{where} \quad d_x \in \mathbb{R}^{m_1} \text{ and } d_y \in \mathbb{R}^{m-m_1} \tag{7}$$

Therefore, Eq. (5) can be rewritten as

$$\begin{aligned} \dot{x} &= f(x) + B_1(x)d_x(t) \\ y &= C(x) + D_1 d_y(t) \end{aligned} \tag{8}$$

where $d_x(t), d_y(t)$ represent the state and the sensor attack vectors, respectively. Different attack strategies are shown in **Table 1** and discussed in Section 1.

Since $p \geq m - m_1$, the system (8) can be partitioned using a nonsingular transformation $M \in \mathbb{R}^{p \times p}$

$$y = M\bar{y} \tag{9}$$

selected so that

$$M^{-1}D_1 = \begin{bmatrix} \mathbf{0}_{(p-(m-m_1)) \times (m-m_1)} \\ \overline{D}_{1_{(m-m_1) \times (m-m_1)}} \end{bmatrix} \tag{10}$$

Taking into account (10), system (8) is reduced to

$$\begin{aligned} \dot{x} &= f(x) + B_1(x)d_x(t) \\ \bar{y}_1 &= C_1(x), \quad \bar{y}_2 = C_2(x) + \overline{D}_1 d_y(t) \end{aligned} \tag{11}$$

where $\bar{y}_1 \in \mathbb{R}^{p_1}$ with $p_1 = p - (m - m_1)$ and $\bar{y}_2 \in \mathbb{R}^{p_2}$ where $p_2 = m - m_1$. Note that the state attack vector $d_x(t)$ is additive and matched to the control input that is embedded in system Eq. (11) already.

| Attack plan | $d_x(t) \neq 0$ | $d_y(t) \neq 0$ | Access to all sensors | Need to know the system model |
|---|---|---|---|---|
| Stealth attack | | √ | | |
| Deception attack | √ | | | |
| Reply attack | √ | √ | √ | |
| Covert attack | √ | √ | | √ |
| False data injection attack | | √ | | √ |

**Table 1.**
*Cyber-attack strategies.*

## 4. Problem formulation

**Assumption (A2):** Attacks are detectable, i.e., the invariant zeros of Eq. (11) are stable.

The problem is to protect the closed loop system (11) from the sensor attack $d_y \in \mathbb{R}^{m-m_1}$ and state/plant attack $d_x(t) \in \mathbb{R}^{m_1}$ by means of designing fixed-gain and adaptive-gain SMOs that allow: (a) reconstructing online the sensor attack $d_y$, the state/plant attack $d_x(t)$, and the plant states $x$ so that

$$\hat{d}_x(t) \to d_x(t), \hat{d}_y(t) \to d_y(t), \hat{x} \to x \qquad (12)$$

as time increases and.

(b) "cleanup" of the plant and sensors so that the dynamics of the CPS under attack (11) approaches,

$$\dot{x}_{clean} = f(\hat{x}) + B_1(\hat{x})\left(d_x(t) - \hat{d}_x(t)\right), \quad y_{clean} = y - D_1\hat{d}_y = C(\hat{x}) + D_1\left(d_y(t) - \hat{d}_y(t)\right). \qquad (13)$$

as time increases, to.

Note that Eq. (13) represents the compensated CPS that converges to CPS without attack as time increases.

## 5. Results: secure state estimation

In this chapter, for the *linearized* case of the system in Eq. (5), two SMOs for state estimation and attack reconstruction are discussed. Two other SMO strategies for nonlinear system (5) are also proposed and investigated.

### 5.1 Attack reconstruction in linear system via filtering by adaptive sliding mode observer

Consider the linearized system in Eq. (5) with $C(x) = Cx$ and $B(x) = B$

$$\dot{x} = Ax + Bd(t), \quad y = Cx + Dd(t) \qquad (14)$$

*5.1.1 System's transformation*

Considering system Eq. (14) and assuming assumption (A1) holds, then as show in [29] there exists a matrix $N \in R^{(n-p)\times n}$ such that the square matrix

$$T_c = \begin{bmatrix} N \\ C \end{bmatrix} \qquad (15)$$

is nonsingular and the change of coordinates $x \mapsto T_c x$ creates, without loss of generality, a new state-space representation $(A', B', C', D)$ where

$$A' = T_c A T_c^{-1}, \quad B' = T_c B, \quad C' = C T_c^{-1} = \begin{bmatrix} 0_{p\times(n-p)} & I_{p\times p} \end{bmatrix} \qquad (16)$$

After the linear changing of coordinate, the CPS Eq. (14) is rewritten as

$$\begin{aligned}\dot{x}_1 &= A_{11}x_1 + A_{12}x_2 + B_1 d \\ \dot{x}_2 &= A_{21}x_1 + A_{22}x_2 + B_2 d \quad \text{where} \quad A' = \begin{bmatrix} A_{11} & A_{12} \\ A_{21} & A_{22} \end{bmatrix}, \quad B' = \begin{bmatrix} B_1 \\ B_2 \end{bmatrix} \\ y &= x_2 + Dd\end{aligned} \quad (17)$$

with $x_1 \in R^{n-p}$, $x_2 \in R^p$, $B_1 \in R^{(n-p) \times m}$, $B_2 \in R^{p \times m}$, $A_{11} \in R^{(n-p) \times (n-p)}$, $A_{12} \in R^{(n-p) \times p}$, $A_{21} \in R^{p \times (n-p)}$, $A_{22} \in R^{p \times p}$. It is well known that $(A, C)$ is observable if and only if $(A_{11}, A_{21})$ is observable [31].

Defining a further change of coordinates $\bar{x}_1 = x_1 + Lx_2$ where $L \in \mathbb{R}^{(n-p) \times p}$ is the design matrix, then the system Eq. (17) can be rewritten as

$$\begin{aligned}\dot{\bar{x}}_1 &= \tilde{A}_{11}\bar{x}_1 + \tilde{A}_{12}x_2 + \tilde{B}_1 d \\ \dot{x}_2 &= \tilde{A}_{21}\bar{x}_1 + \tilde{A}_{22}x_2 + \tilde{B}_2 d\end{aligned}, \quad y = x_2 + Dd \quad (18)$$

where $\tilde{A}_{11} = A_{11} + LA_{21}$, $\tilde{A}_{12} = -A_{11}L + A_{12} - LA_{21}L + LA_{22}$, $\tilde{B}_1 = B_1 + LB_2$, $\tilde{A}_{21} = A_{21}$, $\tilde{A}_{22} = A_{22} - A_{21}L$, $\tilde{B}_2 = B_2$. Since $(A_{11}, A_{21})$ is observable, there exist choices of the matrix $L$ so that the matrix $\tilde{A}_{11} = A_{11} + LA_{21}$ is Hurwitz.

**Assumption (A3):** The attack $d(t)$ and its derivative are norm bounded, i.e., $\|d\| < k_d$ and $\|\dot{d}\| < l_d$ where $k_d, l_d > 0$ and are known.

Since $p > m$, there exists a nonsingular scaling matrix $Q \in R^{p \times p}$ such that

$$QD = \begin{bmatrix} \mathbf{0}_{(p-m) \times m} \\ D_2 \end{bmatrix} \quad (19)$$

where $D_2 \in R^{m \times m}$ is nonsingular. Define $\bar{y}$ as the scaling of the measured outputs $y$ according to $\bar{y} = Qy$. Partition the output of the CPS into unpolluted measurements $\bar{y}_1 \in \mathbb{R}^{p-m}$ and polluted measurements $\bar{y}_2 \in \mathbb{R}^m$ as

$$\bar{y} = \begin{bmatrix} \bar{y}_1 \\ \bar{y}_2 \end{bmatrix} = \begin{bmatrix} Q_1 x_2 \\ Q_2 x_2 + D_2 d \end{bmatrix} = Qx_2 + \begin{bmatrix} \mathbf{0}_{(p-m) \times m} \\ D_2 \end{bmatrix} d \quad (20)$$

Scale state component $x_2$ and define $\bar{x}_2 = Qx_2$. Then Eq. (18) can be rewritten as

$$\begin{aligned}\dot{\bar{x}}_1 &= \overline{A}_{11}\bar{x}_1 + \overline{A}_{12}\bar{x}_2 + \overline{B}_1 d \\ \dot{\bar{x}}_2 &= \overline{A}_{21}\bar{x}_1 + \overline{A}_{22}\bar{x}_2 + \overline{B}_2 d\end{aligned}, \quad \bar{y} = \bar{x}_2 + \begin{bmatrix} 0 \\ D_2 \end{bmatrix} d \quad (21)$$

where $\overline{A}_{11} = \tilde{A}_{11}$, $\overline{A}_{12} = \tilde{A}_{12}Q^{-1}$, $\overline{B}_1 = \tilde{B}_1$, $\overline{A}_{21} = Q\tilde{A}_{21}$, $\overline{A}_{22} = Q\tilde{A}_{22}Q^{-1}$, and $\overline{B}_2 = Q\tilde{B}_2$. Define $\bar{x}_2 = col(\bar{x}_{21}, \bar{x}_{22})$, where $\bar{x}_{21} \in \mathbb{R}^{p-m}$ and $\bar{x}_{22} \in \mathbb{R}^m$. Consequently the system in Eq. (21) can be written in partitioned form as

$$\begin{aligned}\dot{\bar{x}} &= \overline{A}\bar{x} + \overline{B}d \\ \bar{y}_1 &= \overline{C}_1\bar{x}, \quad \bar{y}_2 = \overline{C}_2\bar{x} + D_2 d\end{aligned}, \bar{x} = \begin{bmatrix} \bar{x}_1 \\ \bar{x}_{21} \\ \bar{x}_{22} \end{bmatrix}, \overline{A} = \begin{bmatrix} \overline{A}_{11} & \overline{A}_{12a} & \overline{A}_{12b} \\ \overline{A}_{21a} & \overline{A}_{22a} & \overline{A}_{22b} \\ \overline{A}_{21b} & \overline{A}_{22c} & \overline{A}_{22d} \end{bmatrix}, \overline{B} = \begin{bmatrix} \overline{B}_1 \\ \overline{B}_{21} \\ \overline{B}_{22} \end{bmatrix}$$

$$\overline{C}_1 = \begin{bmatrix} \mathbf{0}_{(p-m) \times (n-p)} & I_{(p-m) \times (p-m)} & \mathbf{0}_{(p-m) \times m} \end{bmatrix}, \overline{C}_2 = \begin{bmatrix} \mathbf{0}_{m \times (n-m)} & I_{m \times m} \end{bmatrix} \quad (22)$$

where $\overline{A}_{11}$ is Hurwitz and the virtual measurement $\bar{y}_1$ presents the protected measurements and $\bar{y}_2$ shows the attacked/corrupted measurements.

*5.1.2 Attack observation*

A SMO is proposed to reconstruct the attack in order to clean up the measurements and states and to allow the use of clean measurement in the control signal.

Define a (sliding mode) observer for the system Eq. (22) as

$$\dot{\bar{z}} = \bar{A}\bar{z} + \bar{G}_1(\bar{y}_1 - \bar{z}_{21}) + \bar{G}_2(\bar{y}_2 - \bar{z}_{22}) - G_n v \qquad (23)$$

where $\bar{z} = col(\bar{z}_1, \bar{z}_{21}, \bar{z}_{22})$ is conformal with the partition of $\bar{x}$ in Eq. (22). In Eq. (23), $v$ is a nonlinear injection signal that depends on $(\bar{y}_2 - \bar{z}_{22})$ and is used to induce a sliding motion in the estimation error space, and

$$\bar{G}_1 = \begin{bmatrix} \bar{A}_{12a} \\ \bar{A}_{22a} - A^s_{22} \\ \mathbf{0}_{m \times (p-m)} \end{bmatrix}, \bar{G}_2 = \begin{bmatrix} \bar{A}_{12b} \\ \bar{A}_{22b} \\ \bar{A}_{22d} - A^s_{33} \end{bmatrix}, G_n = \begin{bmatrix} \mathbf{0}_{(n-p) \times m} \\ \mathbf{0}_{(p-m) \times m} \\ I_{m \times m} \end{bmatrix} \qquad (24)$$

are the gain matrices where $\bar{A}_{12a} \in \mathbb{R}^{(n-p) \times (p-m)}$, $\bar{A}_{22a} \in \mathbb{R}^{(p-m) \times (p-m)}$, $\bar{A}_{12b} \in \mathbb{R}^{(n-p) \times m}$, $\bar{A}_{22b} \in \mathbb{R}^{(p-m) \times m}$, $\bar{A}_{22d} \in \mathbb{R}^{m \times m}$, and the matrices $A^s_{22} \in \mathbb{R}^{(p-m) \times (p-m)}$ and $A^s_{33} \in \mathbb{R}^{m \times m}$ are user-selected Hurwitz matrices, while $A^s_{33}$ is symmetric negative definite. The injection signal $v \in \mathbb{R}^m$ is defined as

$$v = -(\rho + \eta)\frac{\bar{y}_2 - \bar{z}_{22}}{\|\bar{y}_2 - \bar{z}_{22}\|}, \quad \rho, \eta > 0 \qquad (25)$$

where scalar gain $\rho$ will be defined in the sequel, and $\eta$ is a positive design scalar.

**Assumption (A4):** Matrix $(sI - A^*)$ is invertible, where $A^* = \bar{A} - \bar{B}D_2^{-1}\bar{C}_2 - \bar{G}_1\bar{C}_1$.

Defining $\bar{e} = \bar{x} - \bar{z}$, then it follows $\bar{e} = col(\bar{e}_1, \bar{e}_{21}, \bar{e}_{22})$ where $\bar{e}_1 = \bar{x}_1 - \bar{z}_1$, $\bar{e}_{21} = \bar{x}_{21} - \bar{z}_{21}$, $\bar{e}_{22} = \bar{x}_{22} - \bar{z}_{22}$. It follows

$$e_{y_2} = \bar{y}_2 - \bar{z}_{22} = \bar{e}_{22} + D_2 d \qquad (26)$$

and by direct substitution from Eqs. (22) and (23) that

$$\dot{\bar{e}} = \begin{bmatrix} \bar{A}_{11} & 0 & 0 \\ \bar{A}_{21a} & A^s_{22} & 0 \\ \bar{A}_{21b} & \bar{A}_{22c} & A^s_{33} \end{bmatrix} \bar{e} - \begin{bmatrix} \bar{A}_{12b} \\ \bar{A}_{22b} \\ \bar{A}_{22d} - A^s_{33} \end{bmatrix} D_2 d + \begin{bmatrix} \bar{B}_1 \\ \bar{B}_{21} \\ \bar{B}_{22} \end{bmatrix} d + \begin{bmatrix} 0 \\ 0 \\ I_m \end{bmatrix} v \qquad (27)$$

The idea is to force a sliding motion on

$$e_{y_2} = \bar{y}_2 - \bar{z}_{22} = \mathbf{0} \qquad (28)$$

The first main results, based on the SMO with the fixed-gain injection term, is formulated in the following theorem.

**Theorem 1:** Assuming (A3)–(A4) hold and $m_0 > 0$ satisfies the condition

$$\|\phi(t)\| \le m_0 k_d, \quad \phi = [\bar{A}_{21b} \ \bar{A}_{22c}]\bar{e}_{11} - (\bar{A}_{22d} - \bar{B}_{22}D_2^{-1})D_2 d, \quad \bar{e}_{11} = col(\bar{e}_1, \bar{e}_{21}) \qquad (29)$$

Then, as soon as the sliding mode is established in finite time in Eq. (27) on the sliding surface Eq. (28) by means of the injection term Eq. (25) with $\rho = m_0 k_d + \|D_2\|_\infty l_d$, the attack $d$ is asymptotically estimated as

$$\hat{d} = G^*(s)v_{eq} \quad where \quad G^*(s) = C^*(sI - A^*)^{-1}B^*, \; B^* = \begin{bmatrix} \mathbf{0}_{(n-p)\times m} \\ \mathbf{0}_{(p-m)\times m} \\ I_{m\times m} \end{bmatrix}, C^* = \begin{bmatrix} \mathbf{0}_{m\times(n-m)} & -D_2^{-1} \end{bmatrix}$$
(30)

where $v_{eq}$ is the *equivalent* injection term [31] and a close approximation and $\bar{v}_{eq}$ can be obtained in real time by low-pass filtering of the switching signal Eq. (25) [29]. Replacing $v_{eq}$ by $\bar{v}_{eq}$ in Eq. (30) gives

$$\bar{\hat{d}} = G^*(s)\bar{v}_{eq}$$
(31)

Proof of the Theorem 1 is omitted for brevity.

**Remark 1:** The SMO (31) is a dynamic filter that allows reconstructing the time-varying attack $d(t)$. This filter is the main novel feature of the proposed observer.

*5.1.3 Adaptive-gain attack observer design*

In Eq. (29), it was assumed that the perturbation term $\varphi$ is locally norm-bounded and $\rho > 0$ in Eq. (25) is known. In many practical cases, the boundary of attacks is unknown, and the gain of the sliding mode injection term Eq. (25) in the fixed-gain observer in Eq. (23) can be overestimated. The gain overestimation could increase chattering that is difficult to attenuate.

The constant gain $\rho > 0$ can be replaced by an adaptive-gain $\rho(t)$ by applying the *dual layer nested adaptive sliding mode observation algorithm* [32], i.e.,

$$v = -(\rho(t) + \eta)\frac{\bar{y}_2 - \bar{z}_{22}}{\|\bar{y}_2 - \bar{z}_{22}\|}$$
(32)

A sufficient condition to ensure sliding on $e_{y_2} = \mathbf{0}$ in finite time is

$$\rho(t) > \left\| A^s_{33}e_{y_2} + \phi + D_2\dot{d} \right\|$$
(33)

An error signal is defined as

$$\sigma(t) = \rho(t) - \frac{1}{\alpha}\|\bar{v}_{eq}(t)\| - \varepsilon$$
(34)

where the scalars $0 < \alpha < 1$, $\varepsilon > 0$. The adaptation dynamics of $\rho(t)$ in Eq. (32) is defined as [32].

$$\dot{\rho}(t) = -r(t)sign(\sigma(t))$$
(35)

where the time-varying scalar $r(t) > 0$ satisfies an adaptive scheme. It is assumed that $r(t)$ has the structure

$$r(t) = \ell_0 + \ell(t)$$
(36)

where $\ell_0$ is a fixed positive scalar. The evolution of $\ell(t)$ is chosen to satisfy an adaptive law [32]:

$$\dot{\ell}(t) = \begin{cases} \gamma|\sigma(t)| & if\,|\sigma(t)|>\sigma_0 \\ 0 & otherwise \end{cases} \tag{37}$$

where $\gamma > 0$, $\sigma_0 > 0$ are design scalars. The second main results are summarized in Theorem 2 as:

**Theorem 2:** Consider the system in Eq. (27) and

$$a(t) = A_{33}^s e_{y_2} + \phi + D_2\dot{d} \tag{38}$$

and assume that $|a(t)| < a_0$, $|\dot{a}(t)| < a_1$, where $a_0$ and $a_1$ are finite but unknown. A SMO is designed as in Eq. (23) with the *adaptive* injection term in Eqs. (32)–(37). If $\varepsilon > 0$ in (34) is chosen to satisfy

$$\frac{1}{4}\varepsilon^2 > \sigma_0^2 + \frac{1}{\gamma}\left(\frac{qa_1}{\alpha}\right)^2 \tag{39}$$

for any given $\sigma_0$, $q > 1$, and, $0 < \alpha < 1$, then the injection term (32) exploiting the *dual layer adaptive* scheme given by Eqs. (35)–(37) drives $\sigma(t)$ to a domain $|\sigma(t)| < \varepsilon/2$ in finite time and consequently ensures a sliding motion $e_y = 0$ can be reached in finite time and sustained thereafter. The gains $r(t)$ and $\rho(t)$ remain bounded. The sensor attack signal $d(t)$ is reconstructed as in Eq. (30) with the equivalent adaptive injection term $v_{eq}$ or $\bar{v}_{eq}$.

Proof of Theorem 2 is based on the results in [32] and is omitted for brevity.

**Remark 2:** The proposed unit vector injection gain-adaptation algorithm in Eqs. (32)–(37) does not require the knowledge of the boundaries $k_d$, $l_d > 0$ in $\|d\| < k_d$ and $\|\dot{d}\| < l_d$.

### 5.2 State estimation and attack reconstruction in linear systems by using super twisting SMO

Consider the completely observable linearized system Eq. (11) with $C_1(x) = C_1 x$, $C_2(x) = C_2 x$, $B_1(x) = B$, that is,

$$\dot{x} = Ax + B_1 d_x(t), \quad \bar{y}_1 = C_1 x, \quad \bar{y}_2 = C_2 x + \overline{D}_1 d_y(t) \tag{40}$$

where $B_1 \in \mathbb{R}^{n \times m_1}$, $C_1 \in \mathbb{R}^{(p-(m-m_1)) \times n}$, $C_2 \in \mathbb{R}^{(m-m_1) \times n}$.

**Assumption (A5):** The number of uncorrupted/protected measurements is equal or larger than the number of state/plant attack, i.e., $p_1 = p - (m - m_1) \geq m_1$.

The system Eq. (40) is assumed to have an input-output vector relative degree $r = \{r_1, r_2, ..., r_{p_1}\}$, where *relative degree* $r_i$ for $i = 1, 2, ..., p_1$ is defined as follows:

$$\begin{aligned} C_{1i}A^j B_1 &= 0 \quad for \quad all \quad j < r_i - 1 \\ C_{1i}A^{r_i-1}B_1 &\neq 0 \end{aligned} \tag{41}$$

Without loss of generality, it is assumed that $r_1 \leq ... \leq r_{p_1}$.

*5.2.1 Attack observation*

**Assumption (A6):** there exists a full rank matrix.

$$C_a = \begin{bmatrix} C_1 \\ \vdots \\ C_1 A^{r_{\alpha_1}-1} \\ \vdots \\ C_{p_1} \\ \vdots \\ C_{p_1} A^{r_{\alpha_{p_1}}-1} \end{bmatrix} \quad (42)$$

where integers $1 \leq r_{\alpha_i} \leq r_i$ are such that $\text{rank}(C_a B) = \text{rank}(B)$ and $r_{\alpha_i}$ are chosen such that $\sum_{i=1}^{p_1} r_{\alpha_i}$ is minimal.

The following SMO [33] is used to estimate the states of system Eq. (40):

$$\dot{\hat{x}} = A\hat{x} + G_l(y_a - C_a\hat{x}) + G_n v_c(y_a - C_a\hat{x}) \quad (43)$$

where the matrices of appropriate dimensions $G_l$ and $G_n$ are to be designed, and $v_c(.)$ is an injection vector

$$v_c(y_a - C_a\hat{x}) = \begin{cases} -\rho \dfrac{P(y_a - C_a\hat{x})}{\|P(y_a - C_a\hat{x})\|} & \text{if } (y_a - C_a\hat{x}) \neq 0 \\ 0 & \text{otherwise} \end{cases} \quad (44)$$

where $\rho > 0$ is larger than the upper bound of unknown input $d(t)$.

The definition of the symmetric positive definite matrix $P$ can be found in [33]. The auxiliary output $y_a$ is defined by

$$y_a = \begin{bmatrix} y_1 \\ \nu(y_1 - y_1^1) \\ \vdots \\ \nu(\tilde{y}_1^{r_1-1} - y_1^{r_1-1}) \\ \vdots \\ y_{p_1} \\ \vdots \\ \nu(\tilde{y}_{p_1}^{r_{p_1}-1} - y_{p_1}^{r_{p_1}-1}) \end{bmatrix} \quad (45)$$

where the constituent signals in Eq. (45) are given from the continuous second-order sliding mode observer as

$$\begin{aligned} \dot{y}_i^1 &= \nu(y_i - y_i^1) \\ \dot{y}_i^2 &= E_1 \nu(\tilde{y}_i^2 - y_i^2) \\ &\vdots \\ \dot{y}_i^{r_{\alpha_i}-1} &= E_{r_{\alpha_i}-2} \nu(\tilde{y}_i^{r_{\alpha_i}-1} - y_i^{r_{\alpha_i}-1}) \end{aligned} \quad (46)$$

for $1 \leq i \leq p_1$, with

$$\tilde{y}_i^1 = y_i, \quad \tilde{y}_i^j = \nu(\tilde{y}_i^{j-1} - y_i^{j-1}), \quad 2 \leq j \leq r_{\alpha_i} - 1 \quad (47)$$

The scalar function $E_i$ is defined as

$$E_i = 1 \quad if \quad \left|\tilde{y}_j^{i+1} - y_j^{i+1}\right| \leq \varepsilon \, for \, all \, j \leq i, \, else \, E_i = 0 \qquad (48)$$

and the continuous injection term $\nu(.)$ is given by the super twisting algorithm [34]:

$$\begin{aligned} \nu(s) &= \xi(s) + \lambda_s |s|^{1/2} sign(s) \\ \dot{\xi}(s) &= \beta_s sign(s), \quad \lambda_s, \beta_s > 0 \end{aligned} \qquad (49)$$

**Theorem 3:** Assuming the assumptions (A5) and (A6) hold for system Eq. (40), then state/plant attacks are reconstructed as follows:

$$\hat{d}_x = \left((C_a B)^T C_a B\right)^{-1} (C_a B)^T C_a G_n (\nu_c)_{eq} \qquad (50)$$

Proof: Defining the state estimation error as $e = x - \hat{x}$ and the augmented output estimation error $e_y = C_a x - \bar{y}$ with

$$e_y = \left[e_1^1, ..., e_1^{r_{ai}-1}, ..., e_{p_1}^1, ..., e_{p_1}^{r_{ai}-1}\right]^T, \quad y = \left[y_1^1, ..., y_1^{r_{ai}-1}, ..., y_{p_1}^1, ..., y_{p_1}^{r_{ai}-1}\right]^T \qquad (51)$$

then it follows that

$$\dot{e} = x - \dot{\hat{x}} = Ae + B_1 d_x(t) - G_l(y_a - C_a \hat{x}) - G_n \nu_c(y_a - C_a \hat{x}) \qquad (52)$$

By choosing suitable gains $\lambda_s$ and $\beta_s$ in the output injections Eq. (49), then.

$$y_a = C_a x \qquad (53)$$

for all $t > T$ [33]. Then, the error dynamics Eq. (52) is rewritten as

$$\dot{e} = (\overline{A} - G_l C_a)e + \overline{B}_1 d_x(t) - G_n \nu_c(C_a e) \qquad (54)$$

Since $\text{rank}(C_a \overline{B}_1) = \text{rank}(\overline{B}_1)$ and by assumption the invariant zeros of the triple $(A, B, C_a)$ lie in the left half plane, based on the design methodologies in [35], It follows that $e = 0$ is an asymptotically stable equilibrium point of Eq. (52) and dynamics are independent of $d_x(t)$ once a sliding motion on the sliding manifold $s = C_a e = 0$ has been attained. During the sliding mode $\dot{s} = s = 0$, it is

$$\dot{s} = C_a \dot{e} = C_a(\overline{A} - G_l C_a)e + C_a \overline{B}_1 d_x(t) - C_a G_n \nu_c(C_a e) = 0 \qquad (55)$$

as $e \to 0$; then

$$C_a G_n (\nu_c)_{eq} \to C_a \overline{B}_1 d_x(t) \qquad (56)$$

where $(\nu_c)_{eq}$ is the equivalent output error injection required to maintain the system on the sliding manifold. Since $C_a \overline{B}_1$ is full rank, the attack reconstruction is obtained as (50).

According to (A1), $\overline{D}_1$ is full rank; then sensor attacks in Eq. (40) are reconstructed

$$\hat{d}_y(t) = \overline{D}_1^{-1}(\bar{y}_2 - C_2\hat{x}) \tag{57}$$

### 5.3 The state and disturbance observer for nonlinear systems using higher-order sliding mode differentiator

Consider the locally stable system Eq. (11) where $\bar{y}_1$ and $B_1(x)$ are
$\bar{y}_1 = \begin{bmatrix} y_1 & y_2 & ,..., & y_{p_1} \end{bmatrix}^T$, $B = [b_1, b_2, ..., b_{m_1}] \in \mathbb{R}^{n \times m_1}$, $b_i \in \mathbb{R}^n$, $\forall i = 1, ..., m_1$ are smooth vector fields defined on an open $\Omega \subset \mathbb{R}^n$. According to (A5), we consider $p_1 = m_1$ here. The following properties introduced by Isidori [36] are assumed for $x \in \Omega$.

**Assumption (A7):** The system in Eq. (11) is assumed to have vector relative degree $r = \{r_1, r_2, ..., r_{m_1}\}$ and total relative degree $r_t = \sum_{i=1}^{m_1} r_i$, $r_t \leq n$, i.e.,

$$\begin{aligned} L_{b_j}L_f^k y_i(x) &= 0 \quad \forall j = 1, ..., m_1, \; \forall k < r_i - 1, \; \forall i = 1, ..., m_1 \\ L_{b_j}L_f^{r_i-1} y_i(x) &\neq 0 \quad \text{for at least one } 1 \leq j \leq m_1 \end{aligned} \tag{58}$$

**Assumption (A8):** The following Lie derivative matrix is of full rank.

$$L(x) = \begin{bmatrix} L_{b_1}L_f^{r_1-1}y_1 & L_{b_2}L_f^{r_1-1}y_1 & \cdots & L_{b_{m_1}}L_f^{r_1-1}y_1 \\ L_{b_1}L_f^{r_2-1}y_2 & L_{b_2}L_f^{r_2-1}y_2 & \cdots & L_{b_{m_1}}L_f^{r_2-1}y_2 \\ \vdots & \vdots & \vdots & \vdots \\ L_{b_1}L_f^{r_{m_1}-1}y_{m_1} & L_{b_2}L_f^{r_m-1}y_{m_1} & \cdots & L_{b_{m_1}}L_f^{r_{m_1}-1}y_{m_1} \end{bmatrix} \tag{59}$$

**Assumption (A9):** The distribution $\Gamma = \text{span}\{b_1, b_2, ..., b_{m_1}\}$ is involutive [36].

The system given by Eq. (11) with the involutive distribution $\Gamma$ and total relative degree $r_t$ can be rewritten as

$$\begin{aligned} \dot{\delta}_i &= \begin{bmatrix} 0 & 1 & 0 & \cdots & 0 \\ 0 & 0 & 1 & \cdots & 0 \\ \vdots & \vdots & \vdots & \cdots & \vdots \\ 0 & 0 & 0 & 0 & 0 \end{bmatrix}_{r_i \times r_i} \delta_i + \begin{bmatrix} 0 \\ 0 \\ \vdots \\ L_f^{r_i}y_i(x) \end{bmatrix} + \begin{bmatrix} 0 \\ 0 \\ \vdots \\ \sum_{j=1}^{m_1} L_{b_j}L_f^{r_i-1}y_i(x)d(t) \end{bmatrix}, \forall i = 1, ..., m_1 \\ \dot{\gamma} &= g(\delta, \gamma) \end{aligned} \tag{60}$$

where $\delta = \begin{bmatrix} \delta_1 & \delta_2 & \cdots & \delta_{m_1} \end{bmatrix}^T$ and

$$\delta_i = \begin{bmatrix} \delta_{i1} \\ \delta_{i2} \\ \vdots \\ \delta_{ir_1} \end{bmatrix} = \begin{bmatrix} \eta_{i1}(x) \\ \eta_{i2}(x) \\ \vdots \\ \eta_{ir_1}(x) \end{bmatrix} = \begin{bmatrix} y_i(x) \\ L_f y_i(x) \\ \vdots \\ L_f^{r_1-1} y_i(x) \end{bmatrix} \in \mathbb{R}^{r_i} \; \forall i = 1, ..., m_1, \quad \gamma = \begin{bmatrix} \gamma_1 \\ \gamma_2 \\ \vdots \\ \gamma_{n-r} \end{bmatrix} = \begin{bmatrix} \eta_{r+1}(x) \\ \eta_{r+2}(x) \\ \vdots \\ \eta_n(x) \end{bmatrix} \tag{61}$$

With an involutive distribution $\Gamma$ as defined in (A9), it is always possible to identify the variables $\eta_{r+1}(x), ..., \eta_n(x)$ which satisfy

$$L_{b_j}\eta_i(x) = 0 \quad \forall i = r+1, ..., n, \; \forall j = 1, ..., m_1 \tag{62}$$

**Assumption (A10):** The norm-bounded solution of the internal dynamics $\dot{\gamma} = g(\delta, \gamma)$ is assumed to be locally asymptotically stable [29].

If assumption (A9) is satisfied, then it is always possible to find $n - r$ functions $\eta_{r+1}(x), ..., \eta_n(x)$ such that

$$\Psi(x) = col\{\eta_{11}(x), ..., \eta_{1r_1}(x), ..., \eta_{m_11}(x), ..., \eta_{m_1 r_{m_1}}(x), \eta_{r+1}(x), ..., \eta_n(x)\} \in \mathbb{R}^n \quad (63)$$

is a local diffeomorphism in a neighborhood of any point $x \in \overline{\Omega} \subset \Omega \subset \mathbb{R}^n$, i.e.,

$$x = \Psi^{-1}(\delta, \gamma) \quad (64)$$

In order to estimate the derivatives $\delta_{ij}(t)\ \forall i = 1, ..., m_1,\ \forall j = 1, ..., r_i$ of the output. $y_i$ in finite time, higher-order sliding mode differentiators [28] are used here

$$\dot{z}_0^i = v_0^i,\ v_0^i = -\lambda_0^i |z_0^i - y_i(t)|^{(r_i/(r_i+1))} sign(z_0^i - y_i(t)) + z_1^i, \dot{z}_1^i = v_1^i$$
$$\vdots$$
$$\dot{z}_{r_i-1}^i = v_{r_i-1}^i,\ v_{r_i-1}^i = -\lambda_{r_i-1}^i |z_{r_i-1}^i - v_{r_i-2}^i|^{(1/2)} sign(z_{r_i-1}^i - v_{r_i-2}^i) + z_{r_i}^i, \dot{z}_{r_i}^i = -\lambda_{r_i}^i sign(z_{r_i}^i - v_{r_i-1}^i) \quad (65)$$

for $i = 1, ..., m_1$. By construction,

$$\hat{\delta}_1^1 = \hat{\eta}_1^1(x) = z_0^1, ..., \hat{\delta}_{r_1}^1 = \hat{\eta}_{r_1}^1(x) = z_{r_1-1}^1, \dot{\hat{\delta}}_{r_1}^1 = \dot{\hat{\eta}}_{r_1}^1(x) = z_{r_1}^1$$
$$\vdots \quad (66)$$
$$\hat{\delta}_1^{m_1} = \hat{\eta}_1^{m_1}(x) = z_0^{m_1}, ..., \hat{\delta}_{r_{m_1}}^{m_1} = \hat{\eta}_{r_{m_1}}^{m_1}(x) = z_{r_{m_1}-1}^{m_1}, \dot{\hat{\delta}}_{r_1}^{m_1} = \dot{\hat{\eta}}_{r_{m_1}}^{m_1}(x) = z_{r_{m_1}}^1$$

Therefore, the following exact estimates are available in finite time:

$$\hat{\delta}_i = (\hat{\delta}_{i1}, \hat{\delta}_{i2}, ..., \hat{\delta}_{ir_i})^T = (\hat{\eta}_{i1}(\hat{x}), \hat{\eta}_{i2}(\hat{x}), ..., \hat{\eta}_{ir_i}(\hat{x}))^T \in \mathbb{R}^{r_i}$$
$$\forall i = 1, ..., m_1, \quad \hat{\delta} = (\hat{\delta}^1, \hat{\delta}^2, ..., \hat{\delta}^{m_1})^T \in \mathbb{R}^{r_t} \quad (67)$$

Next, integrate Eq. (60) with $\delta$ replaced by $\hat{\delta}$; estimate of internal dynamics is

$$\dot{\hat{\gamma}} = g(\hat{\delta}, \hat{\gamma}) \quad (68)$$

and with some initial condition from the stability domain of the internal dynamics, a asymptotic estimate $\hat{\gamma}$ can be obtained locally

$$\hat{\gamma} = \begin{pmatrix} \hat{\gamma}_1 \\ \hat{\gamma}_2 \\ \vdots \\ \hat{\gamma}_{n-r} \end{pmatrix} = \begin{pmatrix} \hat{\eta}_{r+1}(\hat{x}) \\ \hat{\eta}_{r+2}(\hat{x}) \\ \vdots \\ \hat{\eta}_n(\hat{x}) \end{pmatrix} \quad (69)$$

Therefore, the asymptotic estimate for the mapping (63) is identified as

$$\Psi(\hat{x}) = col\{\hat{\eta}_{11}(\hat{x}), ..., \hat{\eta}_{1r_1}(\hat{x}), ..., \hat{\eta}_{m_11}(\hat{x}), ..., \hat{\eta}_{m_1 r_{m_1}}(\hat{x}), \hat{\eta}_{r+1}(\hat{x}), ..., \hat{\eta}_n(\hat{x})\} \quad (70)$$

asymptotic estimate $\hat{x}$ of the state vector $x$ can be identified via Eqs. (67) and (69)

$$\hat{x} = \Psi^{-1}(\hat{\delta}, \hat{\gamma}) \qquad (71)$$

Since the finite-time exact estimates $\hat{\dot{\delta}}_{ir_i}$ of $\dot{\delta}_{ir_i}$, $\forall i = 1, ..., m_1$ are available via the higher-order sliding mode differentiator, and using the estimates $\hat{\delta}, \hat{\gamma}$ for $\delta, \gamma$, an asymptotic estimate $\hat{d}(t)$ of disturbance $d(t)$ in Eq. (11) is identified as [28].

$$\hat{d}(t) = L^{-1}(\Psi^{-1}(\hat{\delta}, \hat{\gamma})) \left[ \begin{pmatrix} \hat{\dot{\delta}}_{1r_1} \\ \hat{\dot{\delta}}_{2r_2} \\ \vdots \\ \hat{\dot{\delta}}_{m_1 r_{m_1}} \end{pmatrix} - \begin{pmatrix} L_f^{r_1} y_{11}(\Psi^{-1}(\hat{\delta}, \hat{\gamma})) \\ L_f^{r_2} y_{12}(\Psi^{-1}(\hat{\delta}, \hat{\gamma})) \\ \vdots \\ L_f^{r_{m_1}} y_{1m_1}(\Psi^{-1}(\hat{\delta}, \hat{\gamma})) \end{pmatrix} \right] \qquad (72)$$

where $L(\Psi^{-1}(\hat{\delta}, \hat{\gamma})) = \sum_{j=1}^{m_1} L_{b_j} L_f^{r_i - 1} y_{1i}(x)$. Finally, $\hat{x}(t)$ and $\hat{d}(t)$ are obtained from Eqs. (71) and (72).

**Remark 3:** The convergence $\hat{d} \to d$ can be achieved only locally and as time increases due to the local asymptotic stability of the norm-bounded solution of the internal dynamics $\dot{\gamma} = g(\delta, \gamma)$. However convergence will be achieved *in finite time* if the total relative degree $r = n$ and no internal dynamics exist.

Considering Eq. (11) and $\overline{D}_1$ is full rank, sensor attack can be reconstructed as

$$\hat{d}_y(t) = \overline{D}_1^{-1}(\overline{y}_2 - C_2(\hat{x})) \qquad (73)$$

### 5.4 Attack reconstruction in nonlinear system by sparse recovery algorithm

In some applications, there are a limited number of measurements, $p$, and more sources of attack, $m$. Previously, we investigated the cases where $p > m$. Now, consider system (5) with more attacks than measurements, $m > p$.

Notice that a more general format of (5) is considered here where matrix $D$ is a function of $x$ as well.

**Assumption (A11):** Assume that the attack vector $d(t)$ is sparse, meaning that numerous attacks are possible, but the attacks are not coordinated, and only few nonzero attacks happen at the same time.

#### 5.4.1 Sparse recovering algorithm

The problem of recovering an unknown input signal from measurements is well known, as a left invertibility problem, as seen in several works [30, 37], but this problem was only treated in the case where the number of measurements is equal or greater than the number of unknown inputs. The left invertibility problem in the case of fewer measurements than unknown inputs has no solution or more exactly has an infinity of solutions.

In particular, the objective of exact recovery under sparse assumptions denoted for the sake of simplicity as "sparse recovery" (SR) is to find a concise representation of a signal using a few atoms from some specified (over-complete) dictionary,

$$\xi = \Phi \bar{s} + \varepsilon_0 \qquad (74)$$

Secure State Estimation and Attack Reconstruction in Cyber-Physical Systems: Sliding Mode...
DOI: http://dx.doi.org/10.5772/intechopen.88669where $\bar{s} \in \mathbb{R}^N$ are the unknown inputs with no more than $j$ nonzero entries, $\xi \in \mathbb{R}^M$ are the measurements, $\varepsilon_0$ is a measurement noise, and $\Phi \in \mathbb{R}^{M \times N}$ is the dictionary where $M \ll N$.

**Definition 1:** The Restricted Isometry Property (RIP) condition of $j$-order with constant $\varsigma_j \in (0, 1)$ ($\varsigma_j$ is as small as possible for computational reasons) of the matrix $\Phi$ yields

$$(1 - \varsigma_{\bar{s}})\|\bar{s}\|_2^2 \leq \|\Phi\bar{s}\|_2^2 \leq (1 + \varsigma_{\bar{s}})\|\bar{s}\|_2^2 \tag{75}$$

for any $j$ sparse of signal $\bar{s}$. Considering $\Phi_\Gamma$ as the index set of nonzero elements of $\bar{s}$, then Eq. (75) is equivalent to [23]:

$$1 - \varsigma_{\bar{s}} \leq eig(\Phi_\Gamma^T \Phi_\Gamma) \leq 1 + \varsigma_{\bar{s}} \tag{76}$$

where $\Phi_\Gamma$ is the sub-matrix of $\Phi$ with active nodes.

The problem of SR is often cast as an optimization problem that minimizes a cost function constructed by leveraging the observation error term and the sparsity inducing term [37], i.e.,

$$\bar{s}^* = \arg \min_{\bar{s} \in \mathbb{R}^N} \frac{1}{2}\|\xi - \Phi\bar{s}\|_2^2 + \lambda\Theta(\bar{s}) \tag{77}$$

In Eq. (77) the original sparsity term is the quasi norm $|\bar{s}|_0$; but as long as the RIP conditions hold, it can be replaced by $\Theta(\bar{s}) = \|\bar{s}\|_1 \triangleq \sum_i |\bar{s}_i|$. Note that $\lambda > 0$ in Eq. (77) is the balancing parameter and $\bar{s}^*$ is the *critical point*, i.e., the solution of Eq. (74). Typically, for sparse vectors $\bar{s}$ with j-sparsity, where $j$ must be equal or smaller than $\frac{M-1}{2}$ [37], the solution to the SR problem is unique and coincides with the critical point of Eq. (74) providing that RIP condition for $\Phi$ with order $2j$ is verified. In other words, in order to guarantee the existence of a unique solution to the optimization problem Eq. (74), $\Phi$ should satisfy restricted isometry property [37].

Under the sparse assumption of $\bar{s}$ and the fulfillment of the j-RIP condition of the matrix $\Phi$, the estimation algorithm proposed in [37] is

$$\mu\dot{v}(t) = -\lceil v(t) + (\Phi^T\Phi - I_{N\times N})a(t) - \Phi^T\xi \rfloor^\beta, \text{ and } \hat{\bar{s}}(t) = a(t) \tag{78}$$

where $v \in \mathbb{R}^N$ is the state vector, $\hat{\bar{s}}(t)$ represents the estimate of the sparse signal $\bar{s}$ of (74), and $\mu > 0$ is a time-constant determined by the physical properties of the implementing system. $\lceil . \rfloor^\beta = |.|^\beta sign(.)$ and $a(t) = H_\lambda(v)$ where $H_\lambda(.)$ is a continuous soft thresholding function:

$$H_\lambda(v) = \max(|v| - \lambda, 0)\, \text{sgn}\,(v) \tag{79}$$

where $\lambda > 0$ is chosen with respect to the noise and the minimum absolute value of the nonzero terms.

Under Definition 1, the state $v$ of Eq. (78) converges in finite time to its equilibrium point $v^*$, and $\hat{\bar{s}}(t)$ in Eq. (78) converges in finite time to $\hat{s}*$ of Eq. (77).

### 5.4.2 Attack reconstruction

The measured output under attack $y$ of the system Eq. (5) is fed to the input of the low-pass filter that facilitates filtering out the possible measurement noise

$$\dot{z} = \frac{1}{\tau}(-z + C(x) + D(x)d(t)) \tag{80}$$

The filter output $z \in \mathbb{R}^p$ is available. Then, system Eq. (5) with filter Eq. (80) is rewritten as

$$\begin{cases} \dot{\xi} = \eta(\xi) + \Omega d(t) \\ \psi = \overline{C}\xi \end{cases} \tag{81}$$

where $\psi \in \mathbb{R}^p$, and

$$\xi = \begin{bmatrix} z \\ x \end{bmatrix}_{(p+n) \times 1}, \quad \eta(\xi) = \begin{bmatrix} -\frac{1}{\tau}I_{p \times p} & 0 \\ 0 & 0 \end{bmatrix}\begin{bmatrix} z \\ x \end{bmatrix} + \begin{bmatrix} \frac{1}{\tau}C(x) \\ f(x) \end{bmatrix}, \tag{82}$$

$$C = \begin{bmatrix} C_1 & C_2 & \cdots & C_{p+n} \end{bmatrix} = \begin{bmatrix} I_{p \times p} & 0_{p \times n} \end{bmatrix}$$

$$\Omega = \begin{bmatrix} \frac{1}{\tau}D(x) \\ B(x) \end{bmatrix} = \begin{bmatrix} \Omega_1 & \Omega_2 & \cdots & \Omega_m \end{bmatrix}, \quad \Omega_i \in \mathbb{R}^{p+n} \quad \forall i = 1, ..., m$$

If assumption (A2), (A7), and (A9) hold for system Eq. (81), i.e., the relative degree vector of Eq. (81) is $r = \{r_1, r_2, ..., r_p\}$, the distribution $\Gamma = span\{\Omega_1, \Omega_2, ..., \Omega_m\}$ is involutive, and if zero dynamics exist, they are assumed asymptotically stable and may be left alone. Here it is assumed that there are no zero dynamics in system Eq. (81) and it is presented as

$$\dot{\Upsilon}_i = \begin{bmatrix} 0 & 1 & 0 & \cdots & 0 \\ 0 & 0 & 0 & \cdots & 0 \\ \vdots & \vdots & \vdots & \cdots & \vdots \\ 0 & 0 & 0 & 0 & 0 \end{bmatrix}\Upsilon_i + \begin{bmatrix} 0 \\ 0 \\ \vdots \\ L_f^{r_i}\psi_i(\xi) \end{bmatrix} + \begin{bmatrix} 0 \\ 0 \\ \vdots \\ \sum_{j=1}^{m}L_{\Omega_j}L_f^{r_i-1}\psi_i(\xi)d_j \end{bmatrix}, \quad \Upsilon_i = \begin{bmatrix} \Upsilon_1^i(\xi) \\ \Upsilon_2^i(\xi) \\ \vdots \\ \Upsilon_{r_i}^i(\xi) \end{bmatrix} = \begin{bmatrix} \psi_i(\xi) \\ L_f\psi_i(\xi) \\ \vdots \\ L_f^{r_i-1}\psi_i(\xi) \end{bmatrix} \tag{83}$$

for $i = 1, ..., p$, where $\psi_i(\xi)$ is the $i^{th}$ entry of vector $\psi(\xi)$ and satisfies

$$\dot{\Upsilon}_{r_i}^i(\xi) = L_f^{r_i}\psi_i(\xi) + \sum_{j=1}^{m}L_{\Omega_j}L_f^{r_i-1}\psi_i d_j, \quad i = 1, ..., p \tag{84}$$

Then, the following algebraic equation is found from Eq. (84):

$$Z_p = F(\xi)d(t) \tag{85}$$

where $Z_p \in \mathbb{R}^p$, $F(\xi) \in \mathbb{R}^{p \times m}$, and

$$Z_p = \begin{bmatrix} \dot{\Upsilon}_{r_1}^1 \\ \vdots \\ \dot{\Upsilon}_{r_p}^p \end{bmatrix} - \begin{bmatrix} L_f^{r_1}\psi_1(\xi) \\ \vdots \\ L_f^{r_p}\psi_p(\xi) \end{bmatrix}, \quad F(\xi) = \begin{bmatrix} L_{\Omega_1}L_f^{r_1-1}\psi_1 & L_{\Omega_2}L_f^{r_1-1}\psi_1 & \cdots & L_{\Omega_a}L_f^{r_1-1}\psi_1 \\ L_{\Omega_1}L_f^{r_2-1}\psi_2 & L_{\Omega_2}L_f^{r_2-1}\psi_2 & & L_{\Omega_a}L_f^{r_2-1}\psi_2 \\ \vdots & & & \vdots \\ L_{\Omega_1}L_f^{r_p-1}\psi_p & L_{\Omega_2}L_f^{r_m-1}\psi_p & \cdots & L_{\Omega_a}L_f^{r_p-1}\psi_p \end{bmatrix} \tag{86}$$

Finally, filtered system Eq. (5), as it is rewritten in Eq. (85), is in the same form of Eq. (74). Then, sparse recovery algorithm discussed in Section 5.4.1 is applied to Eq. (85) to reconstruct $d(t)$.

**Remark 4:** The derivatives $\dot{\Upsilon}^1_{r_1}, ..., \dot{\Upsilon}^p_{r_p}$ are computed exactly in finite time using higher-order sliding mode differentiators [28] discussed in Eqs. (65) and (66).

## 6. Case study

Consider the mathematical models (1)–(4) of the US Western Electricity Coordinating Council (WECC) power system [8] with three generators and six buses (**Figure 1**) when the sensors of the generator speed deviations from synchronicity are under stealth attack and plant is under deception attack.

**Assumption (A12):** The matrix $L^\theta_{l,l}$ in (3) is nonsingular.

If (A12) holds, then the variable $\theta$ can be rewritten as

$$\theta = \left(L^\theta_{l,l}\right)^{-1}\left(-R^\theta_{l,g}\delta + P_\theta + B_\theta d\right) \tag{87}$$

Substituting (87) into (1), then it follows that

$$\begin{bmatrix}\dot{\delta}\\\dot{\omega}\end{bmatrix} = \begin{bmatrix}0 & I_{p\times p}\\M_g^{-1}\left(-L^\theta_{g,g}+L^\theta_{g,l}(L^\theta_{l,l})^{-1}L^\theta_{l,g}\right) & -M_g^{-1}E_g\end{bmatrix}\begin{bmatrix}\delta\\\omega\end{bmatrix} + \begin{bmatrix}0\\P_{\theta\omega}\end{bmatrix} + \begin{bmatrix}B_\delta\\B_{\theta\omega}\end{bmatrix}d(t), \quad y = C\begin{bmatrix}\delta\\\omega\end{bmatrix} + \begin{bmatrix}D_\delta\\D_\omega\end{bmatrix}d(t)$$

$$P_{\theta\omega} = M_g^{-1}\left(P_\omega - L^\theta_{g,l}(L^\theta_{l,l})^{-1}P_\theta\right), \quad B_{\theta\omega} = M_g^{-1}\left(B_\omega - L^\theta_{g,l}(L^\theta_{l,l})^{-1}B_\theta\right)$$

$$\tag{88}$$

### 6.1 Simulation setup

a. The three sensors of rotor angles, $\delta \in \mathbb{R}^3$, are assumed protected from attack, but the three sensors of the generator speed deviations from synchronicity, $\omega \in \mathbb{R}^3$, are assumed to be attacked.

b. The $B_{1\omega} = I_3$, $B_{1\theta} = 0_{6\times 3}$, $D_\delta = 0_{3\times 6}$ are given, and then Eq. (88) is reduced to

$$\begin{cases}\dot{v} = \varphi_\delta(\delta,\omega),\\\dot{\omega} = \varphi_\omega(\delta,\omega) + P_{\theta\omega} + M_g^{-1}d_x(t)\\y_1 = C_1 v, \quad y_2 = C_2\omega + D_{1\omega}d_y(t)\end{cases}, \text{where } C_1 = C_2 = I_{3\times 3}, D_\omega = \begin{bmatrix}0 & 1 & 2 & 0 & 1 & 1\\1 & 0 & 0 & 2 & 1 & 0\\0 & 0 & 1 & 0 & 1 & 0\end{bmatrix}$$

$$\tag{89}$$

**Remark 5:** $D_{1\omega}$ satisfies RIP condition defined in Eq. (75).

In the first step of attack reconstruction, $d_x(t)$ is estimated by using protected measurements $y_1$ and the SMO described in Section 5.2. It is easy to verify that

$$\begin{array}{l}\overline{C}_{\delta 1}\overline{B} = 0, \quad \overline{C}_{\delta 1}A\overline{B} \neq 0\\\overline{C}_{\delta 2}\overline{B} = 0, \quad \overline{C}_{\delta 2}A\overline{B} \neq 0\\\overline{C}_{\delta 3}\overline{B} = 0, \quad \overline{C}_{\delta 3}A\overline{B} \neq 0\end{array} \quad C_a = \begin{bmatrix}C_1\\C_1A\\C_2\\C_2A\\C_3\\C_3A\end{bmatrix} = \begin{bmatrix}1 & 0 & 0 & 0 & 0 & 0\\0 & 0 & 0 & 1 & 0 & 0\\0 & 1 & 0 & 0 & 0 & 0\\0 & 0 & 0 & 0 & 1 & 0\\0 & 0 & 1 & 0 & 0 & 0\\0 & 0 & 0 & 0 & 0 & 1\end{bmatrix}, y_a = \begin{bmatrix}y_1\\\mu(y_1-\hat{y}_1)\\y_2\\\mu(y_2-\hat{y}_2)\\y_3\\\mu(y_3-\hat{y}_3)\end{bmatrix}$$

$$\tag{90}$$

where $\overline{C}_{\delta i}$ is the *i*th row of $\overline{C}_\delta$. The states of the system, $\hat{\delta}$, $\hat{\omega}$, and plant attacks $\hat{d}_x(t)$ are reconstructed using Eqs. (43) and (50). Then, $\hat{\omega}$ is used in Eq. (89) to find

$$D_\omega d_y(t) = y_2 - \hat{\omega} \quad (91)$$

There are six sources $d_{y1}, ..., d_{y6}$ attacking three measurements $\omega_1, \omega_2, \omega_3$, and at any time, just one out of six attack signals is nonzero. The SR algorithm in Section 5.2 is applied to find $\hat{d}_y(t)$. The following attacks are considered for simulation.

$$\begin{bmatrix} d_{x1} \\ d_{x2} \\ d_{x3} \end{bmatrix} = 1(t-10) \cdot \begin{bmatrix} \sin(0.5t) \\ 1(t) - 1(t-4) + 1(t-8.5) - 1(t-13) + 1(t-17.5) \\ \cos(t) + 0.5\sin(3t) \end{bmatrix}, \quad (92)$$

$$d_y(t) = 1(t-10) \cdot \begin{bmatrix} 0 & 0 & 0 & 0 & \sin(t) & 0 \end{bmatrix}^T$$

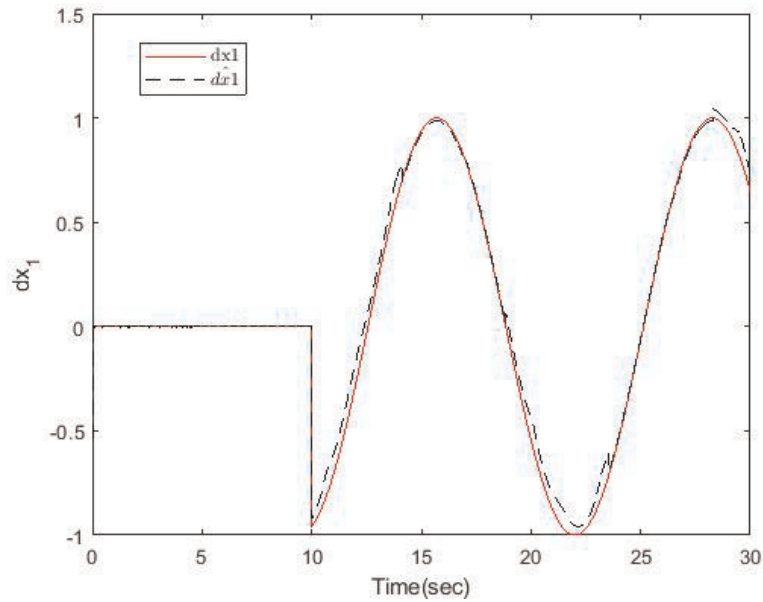

**Figure 4.**
*Plant attack $d_{x_1}$ compared to estimated $\hat{d}_{x_1}$.*

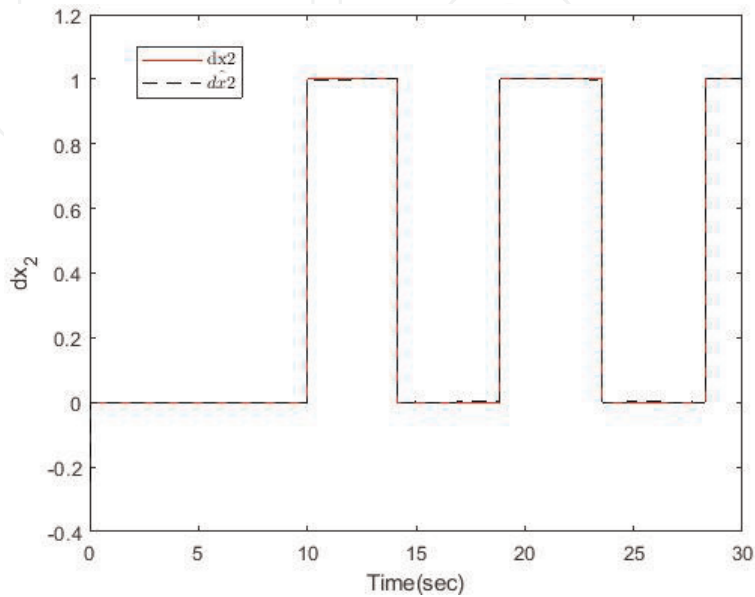

**Figure 5.**
*Plant attack $d_{x_2}$ compared to estimated $\hat{d}_{x_2}$.*

...

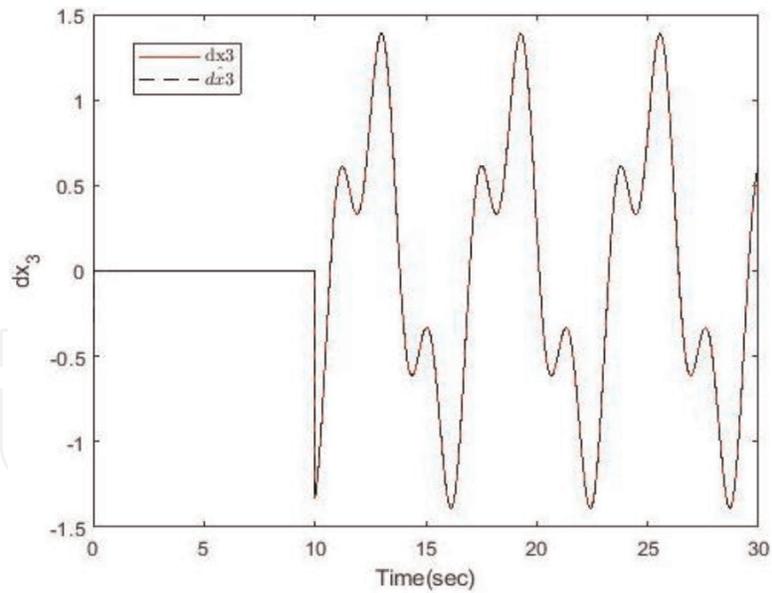

**Figure 6.**
*Plant attack $d_{x_3}$ compared to estimated $\hat{d}_{x_3}$.*

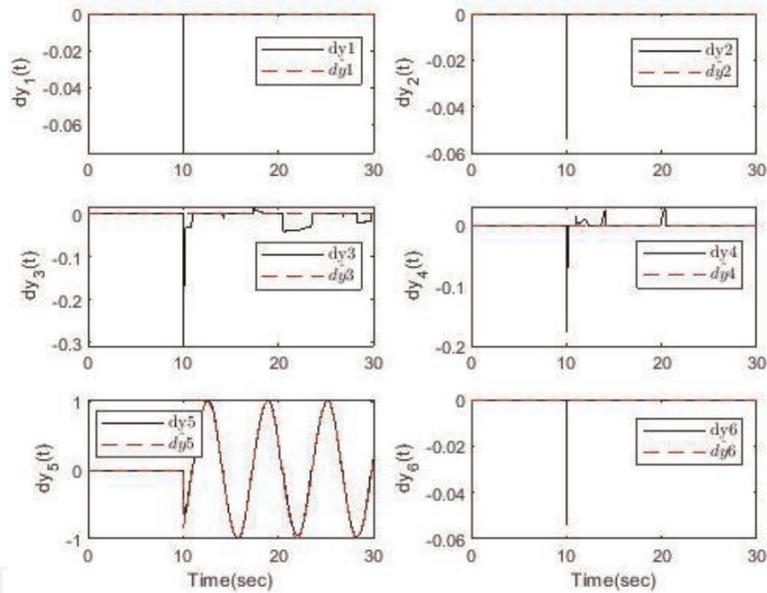

**Figure 7.**
*Sensor attack $d_y$ reconstruction.*

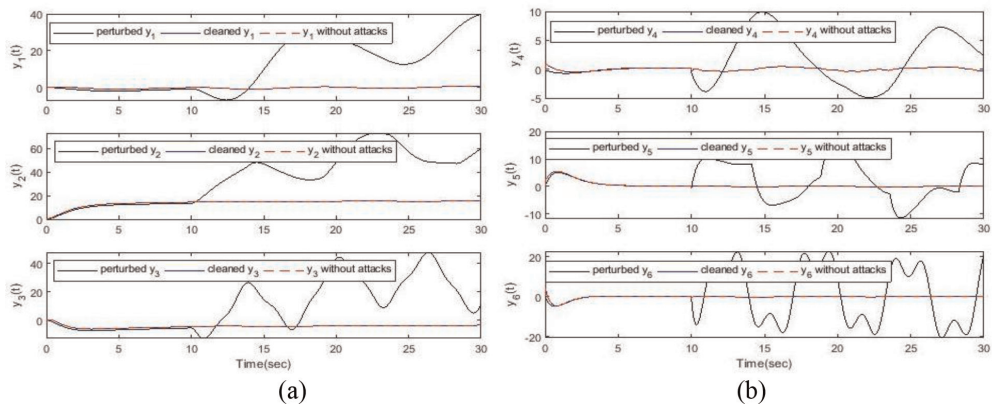

**Figure 8.**
*(a) Corrupted output $y_1, y_2, y_3$ compared with compensated and without any attack output and (b) corrupted output $y_4, y_5, y_6$ compared with compensated and without any attack output.*

Deception attacks $d_{x1}$, $d_{x2}$, and $d_{x3}$ are reconstructed very accurately as shown in **Figures 4–6**. The only nonzero sensor attack is detected and accurately estimated by using the SR algorithm as shown in **Figure 7**. In **Figure 8a** and **8b**, the corrupted system outputs (which are system states in our case) are compared to the "cleaned" outputs that are computed by subtracting the estimated attacks from the corrupted sensors and actuators and to the system outputs when the system is not under attack.

## 7. Conclusion

The critical infrastructures like power grid, water resources, etc. are large interconnected cyber-physical systems whose reliable operation depends critically on their cyber substructure. In this chapter, cyber-physical systems when their sensors and/or states are under attack or experiencing faults are investigated. The sensor and states/plant attacks are reconstructed online by using a fixed-gain and adaptive-gain sliding mode observers. As soon as the attacks are reconstructed, corrupted measurements and states are cleaned from attacks, and the control signal that uses cleaned measurements provides cyber-physical system performance close to the one without attack. The effectiveness of the proposed approach is shown by simulation results of a real electrical power network with sensors under stealth attack and states under deception attacks.

**Author details**

Shamila Nateghi[1]*, Yuri Shtessel[1], Christopher Edwards[2] and Jean-Pierre Barbot[3]

1 The University of Alabama in Huntsville, Huntsville, AL, USA

2 The University of Exeter, Exeter, UK

3 Quartz Laboratory, ENSEA, Cergy-Pontoise, France

*Address all correspondence to: shamila.nateghi.b@gmail.com

**References**

[1] Antsaklis P. Goals and challenges in cyber-physical systems research. IEEE Transactions on Automatic Control. 2014;**59**:3117-3119. DOI: 10.1109/TAC.2014.2363897

[2] Baheti R, Gill H. Cyber-physical systems. The Impact of Control Technology. 2011;**12**:161-166

[3] Conti JP. The day the samba stopped. Engineering and Technology. 2010;**5**:46-47. DOI: 10.1049/et.2010.0410

[4] Karnouskos S. Stuxnet worm impact on industrial cyber-physical system security. In: 37th Annual Conference of the IEEE Industrial Electronics Society 7-10 November 2011; Melbourne: VIC, Australia. 2011. pp. 4490-4494

[5] Farhat A, Cheok CK. Improving adaptive network fuzzy inference system with Levenberg-Marquardt algorithm. In: Annual IEEE International Systems Conference 24-27 April 2017; Montreal: QC, Canada. 2017. pp. 1-6

[6] Farhat A, Hagen K, Cheok KC, Boominathan B. Neuro-fuzzy-based electronic brake system modeling using real time vehicle data. EPiC Series in Computing. 2019;**58**:444-453. DOI: 10.29007/q7pr

[7] Cardenas A, Amin S, Sastry S. Secure control: Towards survivable cyber-physical systems. In: The 28th International Conference on Distributed Computing Systems Workshops. 2008. pp. 495-500

[8] Pasqualetti F, Dörfler F, Bullo F. Control-theoretic methods for cyber-physical security: Geometric principle for optimal cross-layer resilient control systems. IEEE Control Systems Magazine. 2015;**35**:110-127. DOI: 10.1109/MCS.2014.2364725

[9] Mo Y, Sinopoli B. Secure control against replay attacks. In: Proceedings of Allerton Conf. Communications, Control Computing; Monticello: USA. 2009. pp. 911-918

[10] Khazraei A, Kebriaei H, Salmasi RF. Replay attack detection in a multi agent system using stability analysis and loss effective watermarking. In: Annual American Control Conference; Seattle: WA, USA. 2017. pp. 4778-4783. DOI: 10.23919/ACC.2017.7963694

[11] Smith R. A decoupled feedback structure for covertly appropriating network control systems. IFAC Proceedings Volumes. 2011;**44**:90-95. DOI: 10.3182/20110828-6-IT-1002.01721

[12] Mo Y, Sinopoli B. False data injection attacks in control systems. In: Preprints of the 1st Workshop on Secure Control Systems. 2010. pp. 1-6

[13] Gligor VD. A note on denial-of-service in operating systems. IEEE Transactions on Software Engineering. 1984;**SE-10**:320-324. DOI: 10.1109/TSE.1984.5010241

[14] Dan G, Sandberg H. Stealth attacks and protection schemes for state estimators in power systems. In: Proc. IEEE Int. Conf. Smart Grid Communications; USA. 2010. pp. 214-219

[15] Hashemi N, Murguia C, Ruths J. A comparison of stealthy sensor attacks on control systems. In: American Control Conference; Milwaukee: USA. 2018. pp. 973-979

[16] Pasqualetti F, Dorfler F, Bullo F. Attack detection and identification in cyber-physical systems. IEEE Transactions on Automatic Control.

2013;**58**:2715-2729. DOI: 10.1109/TAC.2013.2266831

[17] Jin X, Haddad WM, Yucelen T. An adaptive control architecture for mitigating sensor and actuator attacks in cyber-physical systems. IEEE Transactions on Automatic Control. 2017;**62**:6058-6064. DOI: 10.1109/TAC.2017.2652127.

[18] Nateghi S, Shtessel Y, Barbot JP, Zheng G, Yu L. Cyber-attack reconstruction via sliding mode differentiation and sparse recovery algorithm: Electrical power networks application. In: 15th International Workshop on Variable Structure Systems and Sliding Mode Control; Graz: Austria. 2018. pp. 285-290

[19] Razzaghi P, Khatib EA, Hurmuzlu Y. Nonlinear dynamics and control of an inertially actuated jumper robot. Nonlinear Dynamics. 2019;**97**:161-176. DOI: 10.1007/s11071-019-04963-1

[20] Nateghi S, Shtessel Y. Robust stabilization of linear differential inclusion using adaptive sliding mode control. In: Annual American Control Conference; Milwaukee: USA. 2018. pp. 5327-5331

[21] Navabi M, Mirzaei H. Robust optimal adaptive trajectory tracking control of quadrotor helicopter. Latin American Journal of Solids and Structures. 2017;**14**:1040-1063. DOI: 10.1590/1679-78253595

[22] Razzaghi P, Khatib EA, Bakhtiari S. Sliding mode and SDRE control laws on a tethered satellite system to de-orbit space debris. Advances in Space Research. 2019;**65**:18-27. DOI: 10.1016/j.asr.2019.03.024

[23] Nateghi S, Shtessel Y, Barbot JP, Edwards C. Cyber attack reconstruction of nonlinear systems via higher-order sliding-mode observation and sparse recovery algorithm. In: Conference on Decision and Control; Miami Beach: USA. 2018. pp. 5963-5968

[24] Corradini ML, Cristofaro A. Robust detection and reconstruction of state and sensor attacks for cyber-physical systems using sliding modes. IET Control Theory and Applications. 2017;**11**:1756-1766. DOI: 10.1049/iet-cta.2016.1313

[25] Mousavian S, Valenzuela J, Wang J. A probabilistic risk mitigation model for cyber-attacks to PMU networks. IEEE Transactions on Power Apparatus and Systems. 2015;**30**:156-165. DOI: 10.1109/TPWRS.2014.2320230

[26] Taha A, Qi J, Wang J, Panchal J. Risk mitigation for dynamic state estimation against cyber-attacks and unknown inputs. IEEE Transactions on Smart Grid. 2018;**9**:886-899. DOI: 10.1109/TSG.2016.2570546

[27] Edwards C, Spurgeon SK, Patton RJ. Sliding mode observers for fault detection and isolation. Automatica. 2000;**36**:541-553. DOI: 10.1016/S0005-1098(99)00177-6

[28] Fridman L, Shtessel Y, Edwards C, Yan XG. Higher order sliding mode observer for state pstimation and input reconstruction in nonlinear systems. International Journal of Robust and Nonlinear Control. 2008;**18**:399-412. DOI: 10.1002/rnc.1198

[29] Shtessel Y, Edwards C, Fridman L, Levant A. Sliding Mode Control and Observation. New York: Birkhauser, Springer; 2014

[30] Yu L, Zheng G, Barbot J-P. Dynamic sparse recovery with finite-time convergence. IEEE Transactions on Signal Processing. 2017;**65**:6147-6157. DOI: 10.1109/TSP.2017.2745468

[31] Utkin VI. Sliding Modes in Control Optimization. Berlin: Springer-Verlag; 1992

[32] Edwards C, Shtessel Y. Adaptive continuous higher order sliding mode control. Automatica. 2016;**65**:183-190. DOI: 10.1016/j.automatica.2015.11.038

[33] Floquet T, Edwards C, Spurgeon SK. On sliding mode observers for systems with unknown inputs. International Journal of Adaptive Control and Signal Processing. 2007;**21**:638-656. DOI: 10.1109/VSS.2006.1644520

[34] Levant A. Sliding order and sliding accuracy in sliding mode control. International Journal of Control. 1993;**58**:1247-1263. DOI: 10.1080/00207179308923053

[35] Edwards C, Spurgeon SK. Sliding Mode Control: Theory and Applications. London: Taylor and Francis; 1998. DOI: 10.1201/9781498701822

[36] Isidori A. Nonlinear Control Systems. 3rd ed. Berlin: Springer; 1995. pp. 219-290

[37] Candes E, Tao T. The Dantzig selector: Statistical estimation when p is much larger than n. The Annals of Statistics. 2007;**35**:2313-2351. DOI: 10.1214/009053606000001523